# REVERSE PROXY FRAMEWORK USING SANITIZATION TECHNIQUE FOR INTRUSION PREVENTION IN DATABASE

Vrushali S. Randhe[1], Archana B. Chougule[1], Debajyoti Mukhopadhyay[1]

[1]Department of Information Technology, Maharashtra Institute of Technology, Pune, India
vrushali.randhe@gmail.com, chouguleab@gmail.com,
debajyoti.mukhopadhyay@gmail.com

**Abstract:** With the increasing importance of the internet in our day-to-day life, data security in web application has become very crucial. Ever increasing online and real time transaction services have led to manifold rise in the problems associated with the database security. Attacker uses illegal and unauthorized approaches to hijack the confidential information like username, password and other vital details. Hence the real-time transaction requires security against web based attacks. SQL injection and cross site scripting attack are the most common application layer attack. The SQL injection attacker pass SQL statement through a web application's input fields, URL or hidden parameters and get access to the database or update it. The attacker take a benefit from user provided data in such a way that the user's input is handled as a SQL code. Using this vulnerability an attacker can execute SQL commands directly on the database. SQL injection attacks are most serious threats which take user's input and integrate it into SQL query. Reverse Proxy is a technique which is used to sanitize the users' inputs that may transform into a database attack. In this technique a data redirector program redirects the user's input to the proxy server before it is sent to the application server. At the proxy server, data cleaning algorithm is triggered using a sanitizing application. In this framework we include detection and sanitization of the tainted information being sent to the database and innovate a new prototype.

## INTRODUCTION

The data security is an important aspect to the organizations which are developing web applications. It is a great challenge to the organizations to protect their valuable data against intruders, corruptions and malicious accesses.

The main aim of database security is to provide the protection to the database from unauthorized access. The unauthorized or illegal action may execute the malicious query; because of this vulnerability database security may break. Researchers have developed the Intrusion Detection System (IDS) for enhancing database security against malicious access. Database security can be categorized into external attack and insider attack. In external attack sensitive data is compromised by unauthorized attempts, whereas insider attack is executed by an authorized user with the help of various malicious techniques.

Attacks have increased as the large number of information systems are connected to the internet. SQL injection is one of the most serious security threats for web applications and databases. The attacker can submit a SQL query or command through the web application and can gain access to the underlying confidential information, by bypassing authentication [1] mechanisms, to modify the database and lead to execution of illogical code. This topic focuses on different types of vulnerability in database and the scheme used for detecting the Structure Query Language Injection Attack (SQLIA) and Cross Site Scripting attacks and prevention of the same using reverse proxy from the web applications.

## Webpage Connection

The user (client) and the host (server) communicate through HTTPS in web-based applications. The process starts by entering the selected URL on the browser. The client permits one or more requests for accessing the web contents. The application server accepts the request and processes the same and then responds to the user with the web content. The information can be transmitted by using multiple request/response within a single established connection session. The host performs further processing of the user request and give related response. The web page presented on the browser is the combination of all the related request/response, as shown in Fig. 1. The attacker sends the malicious action through the request or when server sends the response to the client by using certain features of the programming languages or the design of the browser. The attacks occur mainly because of entering, transmitting and combining special grammars, malicious links, text strings

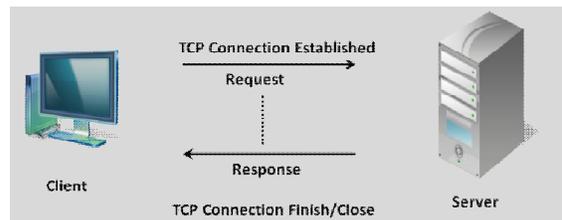

**Fig. 1** HTTPS and Web Page Connection

etc. These attacks usually manage the original web procedure or server response.

**SQL Injection Attack**

SQL Injection Attackers include user data in the SQL query and arbitrary code is added in such a way that a part of the input is understood as SQL code. The attackers [10] always use this malicious SQL codes for input fields of a web form to gain access and update the data. These types of vulnerability allow an attacker to flow commands directly to a web application's underlying database and destroy functionality or confidentiality. There are two techniques of SQLIA i.e. access through input fields and access through URL. In first technique attacker always bypass the authentication of user i.e. password. Attacker can perform this technique through multiple queries, extended stored procedure and 'or' condition etc. In second technique attacker manipulates the query string in URL. This vulnerability can be represented as:

**SELECT * FROM admin WHERE username = 'admin123' AND password = 'secret'**

If in this query the attacker can enter [, , OR 1=1– –] and [ ] instead of [admin123] and [secret], the query would take the form:

**SELECT * FROM admin WHERE username = 'admin123' OR '1=1 – – ' AND password = ' '**

Working of a SQL injection, as shown in the Fig. 2, the condition equation of [1=1– –] and an empty password are mentioned in the query. After checking this condition equation and empty password, query would found a valid row in the table. An attacker may bypass all authentication measures, make changes to the database and gain unlimited access to the sensitive information on the server. Following are the various types of attacks:

**Tautology Attacks.** The tautology attack always uses conditional statements to inject the code. This attack bypasses the authentication of the web page and extracts the important data. The tautological attacker exploits where clause in the query.

**SELECT accounts FROM users WHERE**

**Login ='nil' OR 1=1---AND password = 'nil'**

The conditional statement (OR 1=1) transforms the where clause into tautology. The injected query returns non null value that means the query is successfully executed.

**UNION Attacks.** This technique is combine two queries using UNION keyword. First query is original and second is injected query. In this attack with the help of original query injected query will retrieve the important data of the user.

**SELECT accounts From user WHERE login=' '**

**UNION SELECT credit card WHERE**

**accno=02220 – – AND Password=' '**

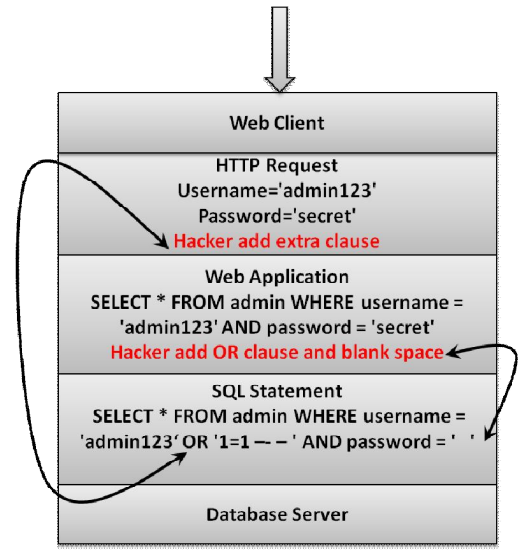

**Fig. 2** Working of SQL Injection

The first/original query returns the null value and second/injected query returns the important data from the "credit card" table.

**Piggybacked Query.** In this type of attack additional query will be added along with original query. The additional query is a injected query which is also called piggy-back onto the original query. Because of additional query database receives multiple SQL queries for execution.

**SELECT accounts FROM user WHERE login='smit' AND pass=''; drop table user – – ' AND pin=231**

After executing the first query, database would find injected/additional query. The result of the above query is to drop the 'user' table and destroy important information.

**Blind Injection.** In this technique, attacker asks true and false question to the server through the web page. If the injected statement evaluates to the true, the web site works normal. If the statement evaluates to false, although there is no descriptive error message, the page differs significantly from the normal functioning page.

**Logical Incorrect Query Attack.** This type of attack gathers important information about the type and structure of the back end database in the web application. When the attacker uses this logical incorrect query, the application server displays error page, which can serve to expose sensitive information about the databases to the attacker.

**SELECT accounts FROM users WHERE login='' AND pass='' AND pin= convert (int,(select top 1 name from sysobjects where xtype='u'))**

The select query extracts the user table and then converts this table name into an integer. Because of this illegal conversion, database gives an error. In this technique, attacker has two useful tricks. First, the attacker can see that the database is an SQL Server database, as the error message explicitly states this fact. Second, the error message reveals the value of the string that caused the type conversion to occur.

## RELATED WORK

Various techniques are proposed in the literature for intrusion detection and prevention. In this section, we brief on some of those techniques.

A random token to SQL keywords and operators are appended in the application code by the SQLrand [3] technique. This approach creates randomized instances of the SQL query language, by randomizing the template query inside the CGI script and the database parser. Here de-randomizing proxy is introduced to allow easy installation of their solutions into existing systems, which converts randomized queries to suitable SQL queries. A proxy server then checks to make sure that all keywords and operators contain this token before sending the query to the database. Because the SQL keywords and operators injected by an attacker would not contain this token, they would be easily recognized as attacks. The shortcoming of this technique is that the token can be easily guessed, making it ineffective, also the approach requires a special proxy server to be deployed.

In SQL Guard and SQL Check [8] and [9] queries are checked at runtime based on a model which is expressed as a grammar that only accepts legal queries. An augmented grammar is used for detection of SQL injection attacks and marking mechanism is used for distinguishing malicious input. SQL Guard examines the structure of the query before and after the addition of user-input based on the model.

CANDID [5] technique supply the programmer-intended input query structure and identify attacks by comparing it against the structure of the actual query issued. Programmer intended queries are mined by dynamically evaluating runs over benign candidate inputs. This method is theoretically well established and is based on inferring intended queries by considering the representative query computed during a program run. CANDID's is very strong tool for detection of SQL injection attacks.

AMNESIA [7] combines static analysis and runtime monitoring. In static analysis it construct model with the help of various type of queries which an application can legally generate at each point of access to the database. Queries are intercepted before they are sent to the database and are checked against the statically built models, in dynamic phase. Queries that violate the model are prevented from accessing to the database. The most important drawback of this technique is that its success is dependent on the accuracy of its static analysis for building query models.

The application SQLIDS [11] produces the intended syntactic structure of SQL statements for security purpose. SQL statements that do not conform to the specifications are considered as security violations and their execution is blocked. Specifications are most important in SQLIDS technique. Specifications are a set of rules which are used to illustrate the expected structure of SQL statement. Generally the application produces the SQL statement. The execution of original SQL statement is expected in the back end database and the specification is created by the syntactic structure Interception of SQL statements. The communication between the web application server and the data server is filtered and the SQL statements that are produced by the application are not transferred directly for execution to the database. This approach detects the injected SQL command by assumption. Therefore, if the intended structure of the expected SQL commands has been explicitly pre-determined, it is possible to detect malicious modifications that alter this structure.

XPATH authentication Technique [2] is used for detection and prevention of SQL Injection attacks in database using web services. To detect and prevent SQLIA with runtime monitoring, without stopping the operation login page is redirected to their checking page. In this technique two modules are used i.e. Active Guard and Service Detector. The first module, Active Guard, is used as vulnerability detector to detect and prevent the susceptibility character and Meta character. Active Guard sends the validated user input to the Service Detector. The user input is sent through the SOAP protocol (Simple Object Access Protocol) to the web service. The user input data is compared with XML_Validator. If the data is matched, the XML_Validator sets a flag value to 1 to Service Detector through the SOAP protocol and valid user can access the web application. If the data is mismatched, the XML_Validator sets a flag value to 0 to Service Detector through the SOAP protocol and invalid user can not access the web application. The second module, Service Detector filtration model is used to validate user input from XPATH_validator where the private data is stored. User input field compare with data existed in XPATH_validator if it is identical then authorized user is allowed for next processing. Web Service Oriented XPATH Authentication Technique does not directly allow to access database in database server.

SQLProb [6] approach is fully modular. In this technique the extracted user input data in the context of the syntactic structure of the query can be evaluated. In this approach access to the source code of the application or the database is not required. Also the system can be easily deployed on the existing enterprise environments and can protect multiple front-end web applications without any modifications. The SQL proxy-based Block (SQLProb) system has four main components i.e. the query collector, the user input extractor, the parse tree generator, and the user input validator. The query collector processes all possible SQL queries during the data collection phase, the user input extractor implements a global pair wise alignment algorithm to identify user input data. The parse tree generator generates the parse tree for the incoming queries. The User Input Validator checks the user input whether it is normal or malicious by using the user input validation algorithm. SQLProb is used in two phases i.e. the data collection phase and the query evaluation phase. In the data collection phase, user input validator collect the queries which cover all the functionalities of the application and save them in storage area. In the query evaluation phase, when a query created by the application is captured by the proxy, the proxy forwards it to the user input extractor and the parse tree generator simultaneously. Experimental results indicated that high detection rate with reasonable performance overhead can be achieved making the system ideal for environments where software or architecture changes is not an economically viable option.

Detecting malicious JavaScript code in Mozilla [13] is a client side technique and uses Mozilla Firefox web browser. This approach focuses on developing an auditing system for JavaScript code execution for detection of anomalies and misuses in SpiderMonkey and Mozilla Firefox web browsers. The auditing system continuously monitors the implementation of JavaScript in these browsers. At the time of execution of JavaScript, various function retracts are registered in the JavaScript engine. These are used when a malicious script tries to access specific information that does not exist in the JavaScript engine. All the calls are intercepted and logged by the auditing system.

The multi agent scanner is used to detect stored cross site scripting attack vulnerabilities [14] in the web applications. Because of the multi agent architecture multiple agents can be able to operate the system independently. The scanner does not require source code of the web application. The system has three modules – web page parser agent, script injector agent and verificator agent. Web page parser agent explores the injection point of stored XSS attack in the web site. The parsing process is analogous to that of the web crawlers and spiders, as it retrieves information from the web pages it visits and spreads through the web site following the hyperlinks it finds. However, two of its characteristics are distinct from the web crawler. First, it follows the hyper-links, targeting the scanned web sites and scrapping all external links. Second, the information recovered is web forms, which are entry points for stored–XSS attacks, as they are the main mechanism supported by the web application for adding new content which is continuously stored by the application. Links containing parameters are not stored as entry points as they are not used to insert new information but generally used to query the database used by the application. They cannot be used for launching stored XSS attacks, but reflected XSS attacks. Script injector agent uses collection of web pages discovered by the web page parser agent. The different attacks get registered through the script agent in performed attack list. The agent injects a set of XSS attack vectors from a known depository into the different input fields of all the injection points. The set of attacks used for the evaluation of this tool are extracted from a depository of XSS attack vectors. These vectors make use of various ways of injecting arbitrary scripts trying to be unobserved by the web application and to be included as valid content in the application. The next module is verificator agent, it uses the performed attacks list of script injector agent and analyses the application. This module corrects the input validation errors. The multi agent scanner scraps with the XSS vulnerabilities in the web application.

Cross-Site Scripting Prevention with Dynamic Data Tainting and Static Analysis [15] proposed a prevention of XSS attack. This technique uses the concept of dynamic data tainting. In this technique sensitive data is marked when scripts are running in the web browser and are dynamically recorded. When malicious data is being transferred to a third party, different actions can be taken, e.g., logging, preventing the transfer and stopping the program with an error. Dynamic taint and static analysis module designed for recording the flow of sensitive information in the web browser and intrusion prevention against cross site scripting attack. When the sensitive information (e.g., user cookies) is about to be transferred to a possible attacker, the user has an option to decide to stop the connection. The Firefox based web crawler is able to simulate user actions. The process allows performing validation of different technique based on automatic browsing.

## PROPOSED FRAMEWORK FOR DATA SANITIZATION

The block diagram of Reverse Proxy Model is illustrated Fig. 3. Reverse proxy server is placed in between the client (request) and the server (response). The client is not aware of the presence of the reverse proxy server. The Reverse proxy server hosts the sanitization application for SQLIA and XSS attacks. This reverse proxy technique is used to sanitize the request from the user or client. The number of reverse proxies can be increased to handle the increased number of requests coming from the client, enabling the system to maintain the least possible response time even at very high loads.

In Reverse Proxy System, clients can't access web servers directly, and the client request must be forwarded by the reverse proxy. There is an intrusion detection and prevention mechanism embedded in the proxy. The basic functional requirements of the web intrusion prevention technique are to receive client requests and server responses, analyse the requests and response messages, decrypt HTTPS messages, forward the safe messages and reject the dangerous ones, and at last log the client requests and detective results. The proxy with intrusion prevention is divided into different parts. First is the request listener, it listens to the ports of reverse proxy server in real time and receives client requests and server responses. Next is the data redirector which extracts the client requests and tokenizes them. All the tokenized parts are saved in a hash table. The tokenized query is then compared with the existing document. The query tokens are transformed into XML format. The attacks are matched with the URI parameters of the HTTP header and the variable pattern is expressed by regular expression.

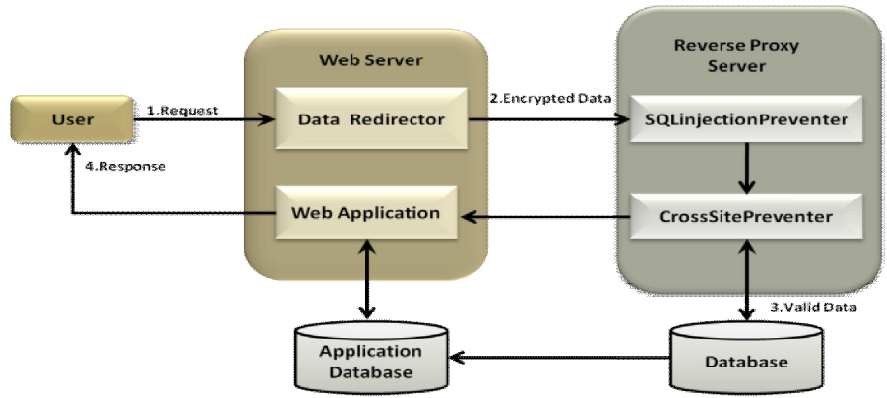

**Fig. 3** Block Diagram of Reverse Proxy Model

The SQL injection preventer and Cross site preventer analyse the client IP addresses. If attacks are encountered more than three times from a particular IP, the IP address is blocked for a pre-determined period and an alert e-mail is sent to the client; otherwise the request will be forwarded to the next module. The blocked IP list can be built not only by the system automatically but can be configured manually by the administrator also. The HTTPS request can be encrypted or decrypted between the client and the web servers. In order to cheat and cross the existing IDS, many hackers use various methods allowed by the web server to encode part of HTTP header or payload, such as URL encoding. Through the encode analysis, it's easy to identify the content of SQL injection and XSS attacks. SQL injection preventer and Cross Site Scripting preventer modules are used to check the intrusion, and if attack action is found, client request is denied.

The log module is used to log the client access requests, discovered attacks, privacy leaks and other system actions. Logs can help the administrators to analyse system performance, track and position the invaders.

Fig. 4 illustrates the block diagram of sanitization application. HTTP request is forwarded to the proxy, which authenticates the request by sending it to the sanitization Application. The sanitization application is combination of SQL preventer and Cross site preventer which is divided into two parts i.e. signature check and MD5 hashing. The sanitization application extracts the URL from the http request and separates the URL and user data. Then the URL and user data is parsed into token [12]. First, Signature check validates the URL by using regular expression and if any invalid character is found in the input query it is marked as malicious and not forwarded to the next module. Only request which is reported as valid by signature check are forwarded to the next module. User data is parsed into tokens. These tokens are checked with the reserve SQL keywords and converted to corresponding MD5.The sanitization application filters the request for invalid tags and encodes it before forwarding to the server. Functionality of signature check and MD5 hashing modules is independent in a sense that the valid request goes to both the modules before getting executed on the web server.

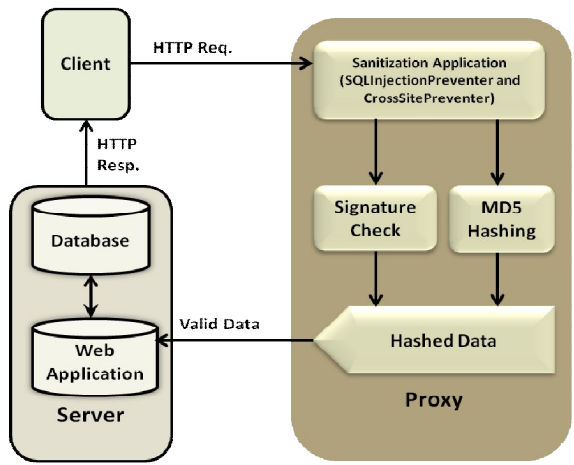

**Fig. 4** Block Diagram of Sanitization Application

## IMPLEMENTATION

The system implements new prototype version of reverse proxy server which prevents the SQL Injection Attacks (SQLIA) and Cross Site Scripting (XSS) attacks. Java is selected for implementation as it is a platform independent language, also during the literature survey we have observed that all types of SQLIA and XSS attacks were successes on the web application which are built using Java. This system uses simple web technologies like HTML, JSP (Java Server Pages), etc. This technique is fully automated. For user inputs we require a web application. Here we are using a banking application. As shown in Fig. 3, following three modules are implemented.

- SQL Injection preventer
- Cross Site Scripting preventer
- Analysis Module

The user sends the input through the login page of web application. The data redirector program, which is installed on the server gateway, gets the user input and redirects the request to the reverse proxy server. At the same time the data redirector encrypts the request into XML format. The reverse proxy server logs the IP addresses of the computers from where the request has originated. The reverse proxy server has the SQL injection preventer module and cross site scripting preventer module installed in it. The SQL injection preventer module validates the request against SQLIA and tokenizes the request. Various signature checks are carried out on the user request, like comments, white spaces, Meta characters, etc. If the special characters are not found in the user request then it is passed on to the cross site scripting preventer module. Cross site scripting preventer module validates the request against XSS attacks by carrying out signature checks through the regular expression. This module prevents the forbidden tags and removes all unwanted and malicious code.

The analysis module checks attacker's activities. If the attacker attacks more than three times consecutively the IP address of the attacker gets blocked for three hour. Also the account holder gets an email notification and the account gets blocked for three hours.

The analysis module can prepare the following reports

(a) Attack's List,
(b) Blocked IP List,
(c) IP Based Analysis and
(d) Web Based Analysis.

In the analysis of Attack's List includes types of attacks, description of attacks, IP address of attackers, browser, URL, and timestamp. The blocked IP analysis includes IP address of attackers and number of attacks from a particular IP. In IP based analysis we observe User-ID, IP address, number of requests and timestamps. The web based analysis displays the browser name and count of attack.

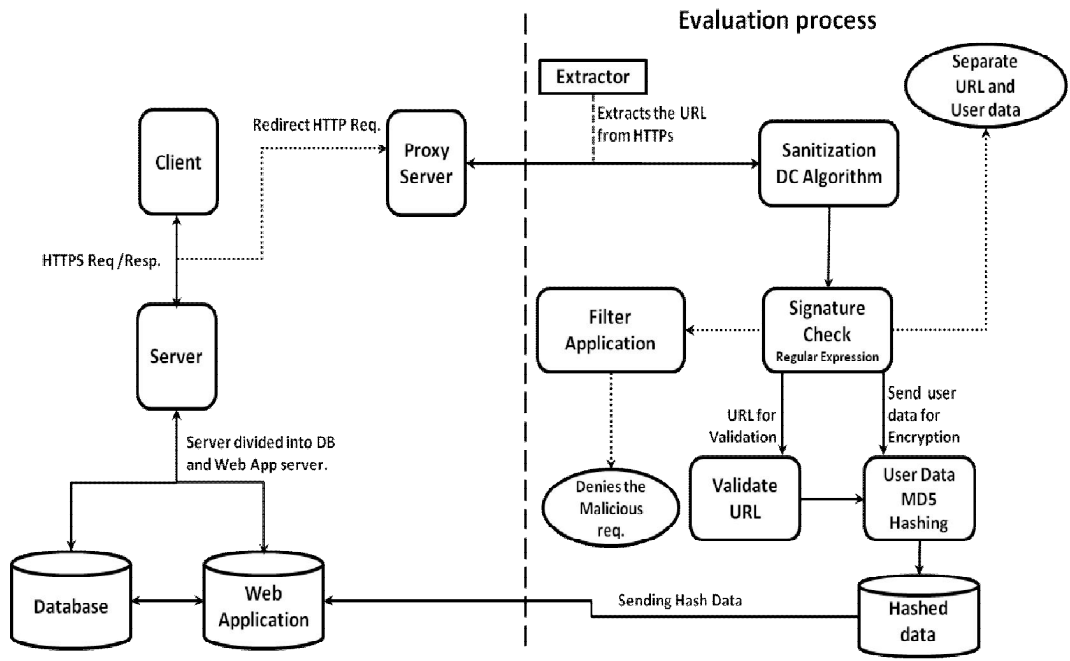

**Fig. 5** The System Architecture

Administrator's Console: This interface creates a table on the basis of analysis done by the analysis module. The table contains following attributes.

- Attacker ID.
- Attackers IP address which linked to view attack and IP base analysis module.
- The login name or password issued to perform the attack.

- Browser details that uses in attack.
- Timestamp when the attack detects.

**Steps for Data Sanitization**

Fig. 5 illustrates the system architecture of reverse proxy. The client sends the request to the server, which is redirected to the reverse proxy server. The sanitizing application in the server extracts the request and separates the URL from HTTP and the user data from the SQL statement. The URL is sent to the signature check where it is tokenized and checked through the regular expression. The user data is encrypted using the MD5 algorithm. The sanitizing application uses SQL injection preventer and Cross Site Scripting preventer algorithm which sends the validated URL and hashed user data to the web application server. If the sanitization application marks the URL request as malicious, the server denies the request. And if the sanitizing application finds the URL request benign, then the hashed value is sent to the database of the web application. When the match is found between the user data and stored hash value in the database, the user gains access to the account, otherwise the access is denied to the user. After completion of the process, the database response is returned to the client and the subsequent request is fetched.

| **Proposed Sanitization Process** |
|---|

**Input**: User data and extracted URL from the HTTP;

**Output**: Access to valid account

**SQL Injection preventer**

1. Take the user request.
2. Parse the user request and store in the hash table and set a key value.
3. The parse or tokenize request checks using signature check through the regular expression.
4. If the tokenize request matches with AND_OR_Integer pattern discard the request. (AND_OR_Integer pattern checks the integer.)
5. If the tokenize request matches with AND_OR_String pattern discard the request. (AND_OR_String pattern checks the string & and/or within single or double quotes.)
6. If the tokenize request matches with SQL Statement Pattern discard the request. (SQL Statement pattern checks the SQL keyword.)
7. Else call the installed Cross Site Scripting preventer program for checking XSS attack

**Cross Site Scripting preventer**

1. Take user request in the form of HTML text.
2. Parse the request and store in the hash table.
3. Check the parse request using signature through regular expression.
4. If the parse request matches with the forbidden tag then discard it.
5. Else remove the tags and encode the URL.
6. If the parse request matches with the attribute tag like href, src and onclick remove that tag and encode the URL and push the tag in the stack.
7. Else remove unknown tags.

| **Steps for Proposed Sanitization Technology** |
|---|

Step 1. Load the configuration in the system
Step 2. The client forwards the request onto the server.
Step 3. The request reaches to the reverse proxy using redirect program.
Step 4. The sanitizing application process the request. Internally, the proxy server extracts the URL direct from HTTP and naturally the user data from the SQL statement. The URL is sent onto the signature check through regular expression.
Step 5. The user data (Using prototype query model) is encrypted using the MD5 hash algorithm. After that check whether is it a query. If yes, send for signature check otherwise send data to the client.
Step 6. The validated URL and hashed user data are sent by the sanitizing application to the server. The filter within the server denies the request in case the sanitizing application had marked the URL request malicious. In the event the URL can be found to become benign, then the hash value is send towards the database of this web application.
Step 7. When the hashed user data and the stored hash value in the database are matched, the user gains access to the account and the proxy server processes for next request. Else the access is denied to the user.

**Extraction of User Data**

As the client sends the HTTP request to the server at the same time SQL statement is extracted from the same request. After the extraction SQL query is tokenized. The tokenized query is compared with the prototype document. A prototype document includes all of the SQL queries that are expected from the web application. The tokenized queries are transformed into XML format [4]. The XSL's pattern matching algorithm is designed to locate the prototype model equivalent to the received Query.

XSL's Pattern Matching: In this pattern matching algorithm, first query is analysed and divided into small elements. The query is first analysed and tokenized as elements. The prototype document stores and maintains the query pertained to the next particular application. The example of the input query is, SELECT * FROM employee WHERE login='admin' AND password= 'XYZ' OR '1=1'.When this query is received, it is converted into XML format using a XML schema.

**Select * from employee where login ='admin' and password='XYZ' OR '1=1'**

**Signature Check**

The signature check handles all possible methods of SQL injection attack through regular expression. The

URL is extracted from the HTTP request and the URL is tokenized. The regular expression checks the small parts of URL. If the regular expression identifies any form of SQL injection signature then the request is marked as malicious else it is marked as benign.

## RUNTIME ANALYSIS

We have evaluated effectiveness of our approach and compared with various other approaches. Table 1 indicates various approaches/preventive techniques including our approach and their effectiveness against different categories of SQL injection and cross site scripting attacks. Our approach works for all approaches listed in the table.

| Query Representation into XML Format |
|---|
| \<Queries\> |
| \<Query id=1\> |
| \<command\> select \</command\> |
| \<project_attribute\>* \</project_attribute\> |
| \<From\> employee \</From\> |
| \<LHS_condition\> login \</LHS_condition\> |
| \<RHS_condition\> string Literal \</RHS_condition\> |
| \<logical_operator\> and \</logical_operator\> |
| \<LHS_condition\> password \</LHS_condition\> |
| \<RHS_condition\> String and Integer Literal \</RHS_condition\> |
| \</Query\> |
| \</Queries\> |

| Approach | SQLIA | | | | | XSS | |
|---|---|---|---|---|---|---|---|
| | Tautology | Logically Incorrect Query | Union Query | Stored Procedure | Piggy-backed query | Stored XSS | Reflected XSS |
| SQLcheck[8] | YES | YES | YES | NO | No | NO | NO |
| SQLRand[3] | YES | NO | YES | NO | YES | NO | NO |
| AMNESIA[7] | YES | YES | YES | NO | YES | NO | NO |
| SQLProb[6] | YES | YES | YES | NO | YES | NO | NO |
| R_ Proxy (Proposed) | YES | YES | YES | YES | YES | YES | YES |

**Table 1** Different Categories of Attacks used in Evaluation of Effectiveness

| Methodology | Change in source code | Detection /Mitigation of Attacks |
|---|---|---|
| SQLCheck[8] | Necessary | Partially Automated |
| SQLRand[3] | Necessary | Fully Automated |
| AMNESIA[7] | Not Necessary | Fully Automated |
| SQLProb[6] | Not Necessary | Fully Automated |
| R_Proxy (Proposed) | Not Necessary | Fully Automated |

Table 2 Analysis of Methodologies Curbing SQLIA

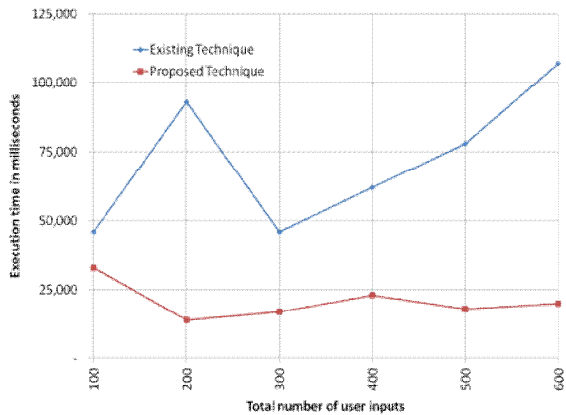

**Fig. 6** Comparison of Time required for Execution

Table 2 indicates comparison between various methodologies with respect to whether changes are required to the source code for intrusion prevention. We also show in the table that our system is fully automated as compared to other partially automated systems for detection of attacks.

We also evaluated execution time of proposed system for varying number of user requests. Table 3 shows that the execution time required by the proposed system is much less than the existing techniques described in [2].

Fig. 6 illustrates comparison between the execution time required by existing technique and the proposed technique. The time required for execution is plotted on 'X' axis in millisecond and total number of user request on 'Y' axis. It can be clearly seen that the proposed technique requires much less time for execution as compared to existing techniques.

| Total User Requests | Execution time in millisecond | |
|---|---|---|
| | Existing Technique | Proposed Technique |
| 100 | 46000 | 33000 |
| 200 | 93000 | 14000 |
| 300 | 46000 | 17000 |
| 400 | 62000 | 23000 |
| 500 | 78000 | 18000 |
| 600 | 107000 | 20000 |

Table 3 Execution Time Comparison

## CONCLUSION

In this paper, various methods for detection and prevention of SQL injection attacks and Cross site scripting attacks are discussed in brief. The goal of Reverse Proxy is to handle security issues. We have proposed a reverse proxy framework for intrusion prevention using sanitization technique. The Reverse proxy model is based on sanitization technique for discovering SQLIA and XSS attacks. The novel system with intrusion prevention proxy is very effective in detecting and preventing the attacks from penetrating the web application. With the help of SQL injection preventer and Cross site scripting preventer, we are able to protect web application against SQLIA and XSS attack. Reverse proxy technique is capable of preventing the malicious script without making any changes to the source code of the application. The analysis module make annotations the attacker history like IP address, timestamp, browser details, URL, etc. The analysis helps to make the system more efficient and flexible. Reverse proxy makes communication between client and web server more secure and efficient. Thus, the user's confidential information is protected while they perform any online transaction through internet.